# Plasmon Engineering in Intercalated 2H-TaS$_2$

Luigi Camerano,[1,2,*] Laura Martella,[1] Lorenzo Battaglia,[1] Federico Giannessi,[1,2] Filippo Camilli,[1] Luca Lozzi,[1] Polina M. Sheverdyaeva,[3] Paolo Moras,[3] Luca Ottaviano,[1,2] Gianni Profeta,[1,2] and Federico Bisti[1]

[1]*Department of Physical and Chemical Sciences, University of L'Aquila, Via Vetoio, 67100 L'Aquila, Italy*
[2]*CNR-SPIN L'Aquila, Via Vetoio, 67100 L'Aquila, Italy*
[3]*CNR-Istituto di Struttura della Materia (CNR-ISM), Strada Statale 14, km 163.5, 34149 Trieste, Italy*

Plasmons in low-dimensional materials provide a powerful platform for nanoscale control of light–matter interactions, yet strategies to tailor their coherence and dissipation remain limited. Here, we demonstrate that transition-metal intercalation offers a fundamentally distinct route to engineer plasmonic response in layered materials. By combining high-resolution core-level photoemission spectroscopy with first-principles calculations, we show that Fe and Co intercalation in 2H-TaS$_2$ does not act as conventional electron doping, but instead reshapes the low-energy electronic structure through orbital hybridization and structural reconstruction. This process introduces a dense continuum of low-energy excitations that efficiently damp and ultimately suppress the plasmon mode. First-principle calculations of the energy-loss function reveal a transition from a well-defined collective excitation to an overdamped response, signaling the breakdown of coherent charge dynamics. Our results establish intercalation as a chemically controlled pathway to tune plasmon losses and dielectric response in quantum van der Waals materials, providing a new design principle for plasmonic and optoelectronic functionalities at the nanoscale.

Collective electronic excitations encode fundamental properties of quantum materials, reflecting the intricate interplay between electronic structure, dimensionality, and many-body interactions [1, 2]. Among these, plasmons, coherent oscillations of the charge density, are uniquely sensitive to low-energy electronic states and dynamical screening [3–5]. Their tunability under external stimuli has positioned plasmons at the forefront of condensed-matter physics and nanophotonics, with broad implications for optoelectronics, sensing, and quantum technologies [6–10]. In low-dimensional materials, where Coulomb interactions and band topology are enhanced, plasmon modes are extremely sensitive to subtle changes in crystal structure and orbital character, offering a direct window into emergent electronic phenomena [11–14]. Layered transition-metal dichalcogenides (TMDs) are an ideal platform to investigate and control these collective excitations [15–18], thanks to their van der Waals structure, which allows chemical manipulation via gating [19], substitutional doping [20], and intercalation [21], accessing a wide range of correlated and topological phases. Among these, transition-metal intercalation is particularly powerful, enabling long-range magnetic order [22–24], spin-textured phases [25–28], and tunable electronic properties [29, 30]. Yet, despite this versatility, how intercalation modifies plasmon excitations and the overall dynamical electronic response remains largely unexplored. Here, we address this challenge by studying plasmon excitations in 2H-TaS$_2$, a prototypical layered metal with intertwined metallicity, charge-density-wave (CDW) order, and hyperbolic light dispersion [31–35]. In this system, electron-energy loss spectroscopy (EELS) has revealed a plasmon mode with negative momentum dispersion [15], reflecting the low-energy electronic structure governed by isolated metallic bands [18, 35–37]. Such collective excitations leave measurable fingerprints in core-level spectra, arising from both many-body screening and inelastic loss processes. Early Ta-$4f$ measurements indeed revealed complex, multi-component lineshapes [38–40], suggesting a rich interplay between electronic structure and collective modes. However, only a quantitative lineshape description can directly connect these spectral features to plasmon dispersion and its evolution upon intercalation. By combining high-resolution synchrotron-radiation-based core-level spectroscopy and laboratory X-ray photoelectron spectroscopy (XPS) with first-principles calculations, we investigate plasmon excitations in 2H-TaS$_2$ and its Fe and Co-intercalated compounds, namely the Ising ferromagnet Fe$_{1/3}$TaS$_2$ [41–44], and the non-coplanar antiferromagnet Co$_{1/3}$TaS$_2$ [45–47]. By carefully modeling the core-level lineshape and explicitly accounting for low-energy excitations arising from the electronic density of states (DoS) calculated from first-principles Density Functional Theory (DFT), we show that plasmonic features strongly shape the spectra of 2H-TaS$_2$. Upon intercalation, these low-energy excitations are strongly suppressed in both Fe$_{1/3}$TaS$_2$ and Co$_{1/3}$TaS$_2$, reflecting the emergence of additional damping channels driven by cooperative structural reconstruction and orbital hybridization rather than simple electron doping. These results demonstrate that transition-metal intercalation provides a chemically controlled route to engineer plasmon losses and, more broadly, the collective electronic response in layered materials.

Bulk 2H-TaS$_2$ is formed by stacked 1H-TaS$_2$ monolayers (MLs) in an ABA sequence related by a glide-mirror symmetry. The 1H-TaS$_2$ ML consists of a hexagonal lattice of Ta atoms in trigonal prismatic coordination with the chalcogen [48]. Fe and Co intercalation in the interlayer spacing of metallic 2H-TaS$_2$ is ordered into a $\sqrt{3} \times \sqrt{3}R30°$ reconstruction and induce magnetism in the system as shown in Fig. 1a-b. Our Low Energy Electron Diffraction (LEED) patterns is accordingly composed by the $1 \times 1$ spots and the additional periodicity compatible with $\sqrt{3} \times \sqrt{3}R30°$ reconstruction (see the orange and green reciprocal space lattice vectors for Fe and Co, respectively in Fig. 1c-d).

* email: luigi_camerano@outlook.it

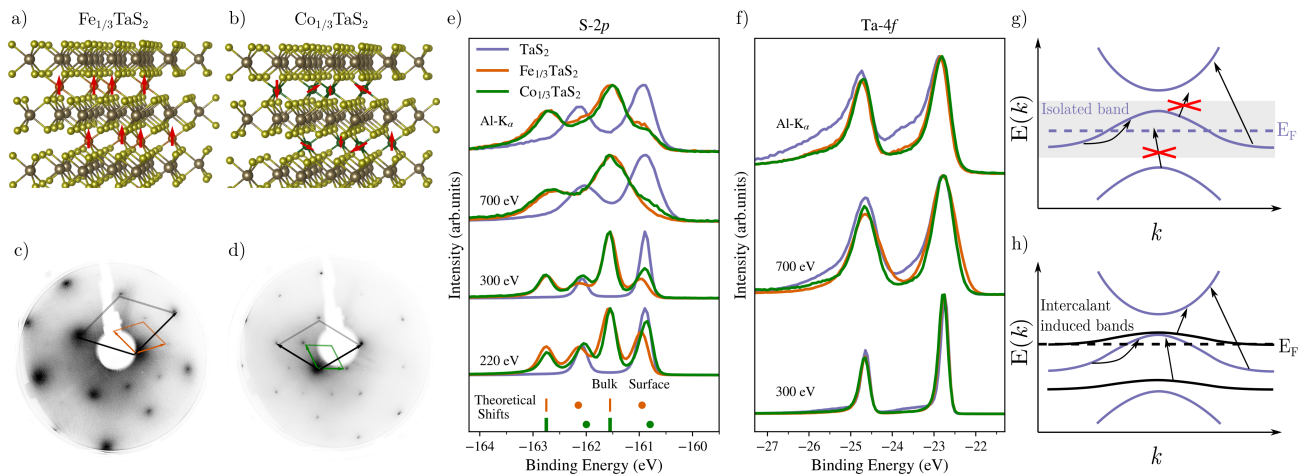

FIG. 1. a) Crystal structure of ferromagnetic Fe$_{1/3}$TaS$_2$ and b) non-collinear triple-Q state of Co$_{1/3}$TaS$_2$ with c-axis aligned along the $z$-direction. c) and d) LEED spots highlighting the $\sqrt{3} \times \sqrt{3}R30°$ reconstruction in both Fe$_{1/3}$TaS$_2$ and Co$_{1/3}$TaS$_2$, respectively. e) S-2$p$ core-level lineshape for different photon energies and different compounds. Photon-dependent spectra were acquired at T = 20 K at VUV beamline (Elettra) while Al-K$_\alpha$ spectra (h$\nu$ = 1486.7 eV) were acquired at room temperature. In the bottom of the figure we report theoretical bulk-surface core-level shifts calculated from initial state theory. f) Photon-energy dependent Ta-4$f$ core-level for 2H-TaS$_2$, Fe$_{1/3}$TaS$_2$ and Co$_{1/3}$TaS$_2$ acquired at h$\nu$ = 300 eV and h$\nu$ = 700 eV. Sketch of the interband transition facilitated by intercalant induced band from 2H-TaS$_2$ g) into Fe/Co$_{1/3}$TaS$_2$ h). The black arrows indicate allowed electronic transitions from occupied to unoccupied states, while the red crosses mark forbidden transitions.

Beyond structural reconstruction, intercalation strongly alters the electronic properties of 2H-TaS$_2$. Figure 1e shows the photon-energy–dependent S-2$p$ core-level spectra aligned to the Fermi level as determined from the valence band (see SI for the details, Figs. S1-S2). Using Al-K$_\alpha$ (h$\nu$ = 1486.7 eV), pristine 2H-TaS$_2$ exhibits a single S-2$p$ main component centered at $\sim -160.9$ eV, with no additional features emerging upon varying the photon energy. In contrast, Fe$_{1/3}$TaS$_2$ and Co$_{1/3}$TaS$_2$ show a chemical shift of the main sulfur peak relative to the pristine compound, together with the appearance of a low-binding-energy shoulder that evolves into a well-resolved peak at lower photon energies. The intensity of this additional component increases with decreasing photon energy, consistent with a surface contribution associated with a sulfur-terminated layer (see SI for angle-dependent XPS, in particular Figs. S5-S6). Notably, the binding energy of the S-2$p$ peak in pristine 2H-TaS$_2$ closely matches that of this surface component in the intercalated systems. Our first-principles slab calculations within the initial-state approximation reproduce this bulk–surface splitting (0.6 eV for Fe$_{1/3}$TaS$_2$ and 0.75 eV for Co$_{1/3}$TaS$_2$, see the bottom of Fig. 1e), confirming that intercalation delocalizes charge around S atoms, reducing its nucleus electronic screening during photoemission process. The combined bulk shift and surface feature provide a clear spectroscopic fingerprint of intercalant–sulfur bonding. These modifications of the S chemical environment likely influence magnetic interactions, naturally explaining deviations from a purely RKKY-mediated exchange, where magnetic coupling arises from the interplay of RKKY and chalcogen-mediated superexchange [49].

In Fig. 1f we report the photon-energy-dependent Ta-4$f$ core-level spectra, aligned to the Fermi level as determined from the valence band (see SI for the details). We first observe that, despite the doping of the valence band induced by intercalation as reported in Refs. [24, 50–52], the Ta-4$f$ core-level binding energy does not exhibit any appreciable shift among the three compounds. This indicates that the changes in carrier concentration do not significantly modify the local electrostatic potential at the Ta site, consistent with efficient metallic screening (this is in contrast with prior studies on these compounds [53]). Beyond the two sharp peaks corresponding to Ta-4$f_{7/2}$ and Ta-4$f_{5/2}$, we observe, particularly in 2H-TaS$_2$, a broad spectral feature in the high-binding-energy tail. Its intensity increases at higher photon energy and it gets broader in the intercalated compounds. The enhancement of this feature at higher photon energies points to its origin as an extrinsic loss process, arising from inelastic scattering of the photoelectron before escaping the sample (see SI for a detailed discussion excluding con-tamination effects). Indeed, increasing photon energy leads to a larger inelastic mean free path, resulting in greater bulk sensitivity and a higher probability for photoelectrons to undergo inelastic scattering events during their propagation to the surface [54]. Such a feature can also be traced in the S-2$p$ core level, although it is less pronounced (see Fig. S5 in SI). Moreover, EELS measurements on pristine 2H-TaS$_2$ confirm the presence of a plasmon resonance at $\sim 1$ eV [15], and previous analyses of the core-level spectra required additional components to accurately reproduce the lineshape [38, 39]. The origin of this plasmon excitation in 2H-TaS$_2$ can be traced to the isolated metallic band characteristic of 2H metallic phases of TMDs [18, 35–37], as illustrated in the sketch in Fig. 1g. The observed suppression of this signal in core-level spectroscopy can be understood as a consequence of the additional electronic states introduced by the intercalated transition metal as recently measured in angle-resolved photoemission spectroscopy (ARPES) [24, 51, 52], which open new low-energy absorption channels (see sketch in Fig. 1h).

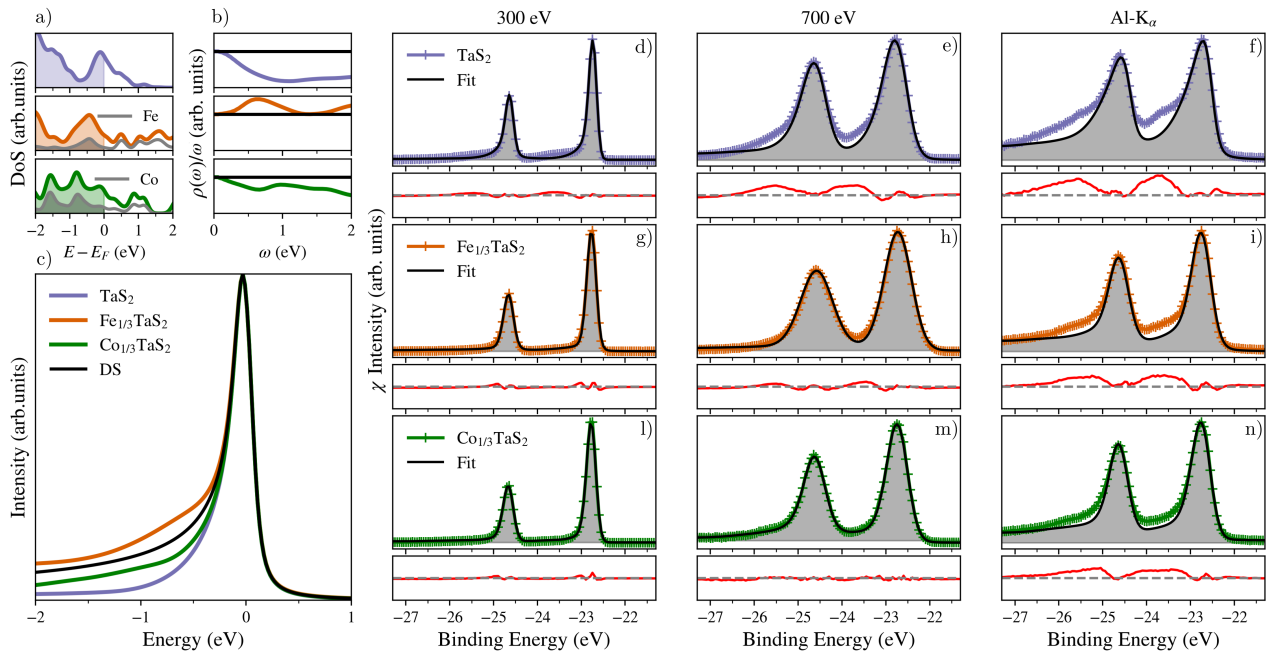

FIG. 2. a) We report the Density of State (DoS) and b) $\rho(\omega)/\omega$ (see main text) in 2H-TaS$_2$, Fe$_{1/3}$TaS$_2$ and Co$_{1/3}$TaS$_2$, respectively. In b) we also reported the $\rho(\omega)/\omega$ for a Doniach-Sunjach (DS) lineshape corresponding to a flat DoS. In c) we report the material dependent core-level lineshape definend by Eq. (1). d), e) and f) Photon-energy dependent core-level fit with the lineshape defined in Eq. 1 for 2H-TaS$_2$. Below we report the difference $\chi$ between the acquired data and the fitting function. The energy scale of this plot is fixed by the values in Fig. 2. In l)-m)-n) and h)-i)-l) the same quantity for Fe$_{1/3}$TaS$_2$ and Co$_{1/3}$TaS$_2$ respectively.

To support this interpretation, we calculated the density of states (DoS), shown in Figs. 2a. Indeed, we find that Fe intercalation induce $n$-doping the Ta-$d^{z^2}$ bands at the Fermi level [24, 34] and introduces additional low-energy states arising from Fe-$d$ and Ta-$d$ hybridization. In contrast, Co intercalation produces a stronger modification of the pristine DoS. This trend is consistent with ARPES measurements [24, 34, 50], which also report an increased density of states near $E_F$ in Co-intercalated TaS$_2$ with respect to Fe intercalated counterpart. Moreover, the calculation of the DoS allows us to quantitatively isolate the extrinsic loss contribution by modelling the core-level lineshape from first-principle calculation. Indeed, modifications of the DoS near the Fermi level in metals affect the shape of the many-electron singularity observed in X-ray photoemission [55, 56]. The widely used Doniach–Šunjić (DS) asymmetric lineshape can be generalized to include the influence of the electronic DoS by expressing the core-level intensity in the time domain [57–60]:

$$I(E) \sim \int_{-\infty}^{+\infty} e^{i(E-E_0)t - \lambda|t| - \frac{\sigma^2 t^2}{2} - g(it)} dt$$
$$g(\tau) = \alpha \int_0^\infty \frac{\rho(E)\left(1 - e^{E\tau}\right)}{E^2} dE \qquad (1)$$

where $E_0$ sets the peak position, $\lambda$ and $\sigma$ are the Lorentzian and Gaussian widths, respectively. The function $g(\tau)$ encodes information about the electronic structure near the Fermi level through $\rho(\omega) = JDoS(\omega)/JDoS'(0)$, with $JDoS$ denoting the joint DoS and the prime indicating an energy derivative. The dimensionless parameter $\alpha$ reflects the strength of the many-body interactions, i.e., the influence of the low-energy $JDoS$ on the lineshape [55, 57]. In Fig. 2b we report $\rho(\omega)/\omega$ for the different materials, together with the reference DS $\rho$ obtained by assuming a constant DoS. Since the $JDoS$ measures the number of available electronic states at a given energy, the shape of $\rho(\omega)/\omega$ reflects the low-energy electronic structure near the Fermi level. For pristine 2H-TaS$_2$, the Fermi energy lies near the maximum of the DoS, resulting in a decrease of $\rho(\omega)/\omega$ as one moves away from zero energy. In contrast, Fe$_{1/3}$TaS$_2$ shows Fe-induced states both above and below the Fermi level, which increase the number of available states away from $E_F$, effectively opening additional channels for core-level excitations. Co$_{1/3}$TaS$_2$ exhibits an intermediate behavior, as Co states contribute significantly near the Fermi level while also enhancing the low-energy DoS, leading to a more modest variation of $\rho(\omega)/\omega$ with energy. From Eq. (1), it follows that the shape of $\rho(\omega)/\omega$ directly influences the core-level lineshape, producing distinct asymmetric tails for the three compounds. This is illustrated in Fig. 2c, where the core-level spectra are computed using Eq. (1) with $\alpha = 0.2$.
Therefore, in Fig. 2d–e–f, 2g–h–i, and 2l–m–n we report the results of the fitting procedure using the calculated lineshape for 2H-TaS$_2$, Fe$_{1/3}$TaS$_2$, and Co$_{1/3}$TaS$_2$, respectively, at different photon energies. The fitted values of the asymmetry parameter $\alpha$ are reported in Table I. For all compounds, $\alpha$ shows an overall increase with photon energy, consistent with the progressive validity of the sudden approximation. Indeed, as discussed by Gadzuk and Šunjić [61], higher photoelectron kinetic energies lead to a faster creation of the core hole, driving the lineshape toward the asymmetric limit. The larger fitted values of $\alpha$ from 2H-TaS$_2$ to Co$_{1/3}$TaS$_2$ and Fe$_{1/3}$TaS$_2$ indicate an increasing density of states near the Fermi level and, consequently, a larger $JDoS$. This trend is consistent with the calculated DoS shown in Fig. 2a, linking the asymmetry parameter $\alpha$ to the number of available electronic states at $E_F$. A quantitative analysis of the residual $\chi$, defined as the difference between the experimental spectra and the fitting function, further highlights a feature at $\sim 1$ eV from the main peak in 2H-TaS$_2$ (see Fig. 2b–c–d), which cannot be reproduced by the core-level lineshape modeling, even when including electronic structure effects. Applying the same analysis to Fe$_{1/3}$TaS$_2$ and Co$_{1/3}$TaS$_2$ reveals a marked suppression of this feature. In Fe$_{1/3}$TaS$_2$, a residual signal is still visible at $h\nu = 700$ eV and with the Al-K$_\alpha$ lamp source (see Fig. 2e–f–g). In Co$_{1/3}$TaS$_2$, the feature is further suppressed and remains barely detectable, broad and weak, only in the Al-K$_\alpha$ spectra (see Fig. 2h–i–l). Overall, the Ta-4$f$ core-level analysis highlights a pronounced suppression of the extrinsic plasmon-loss signal upon intercalation (in Figs. S3-S4-S5-S6 in SI we propose a fit with an asymmetric lineshape as suggested in Refs. [54, 62–64]). At the same time, the spectra indicate efficient metallic screening and a modification of the low-energy electronic structure, consistent with a reduced DoS near the Fermi level.

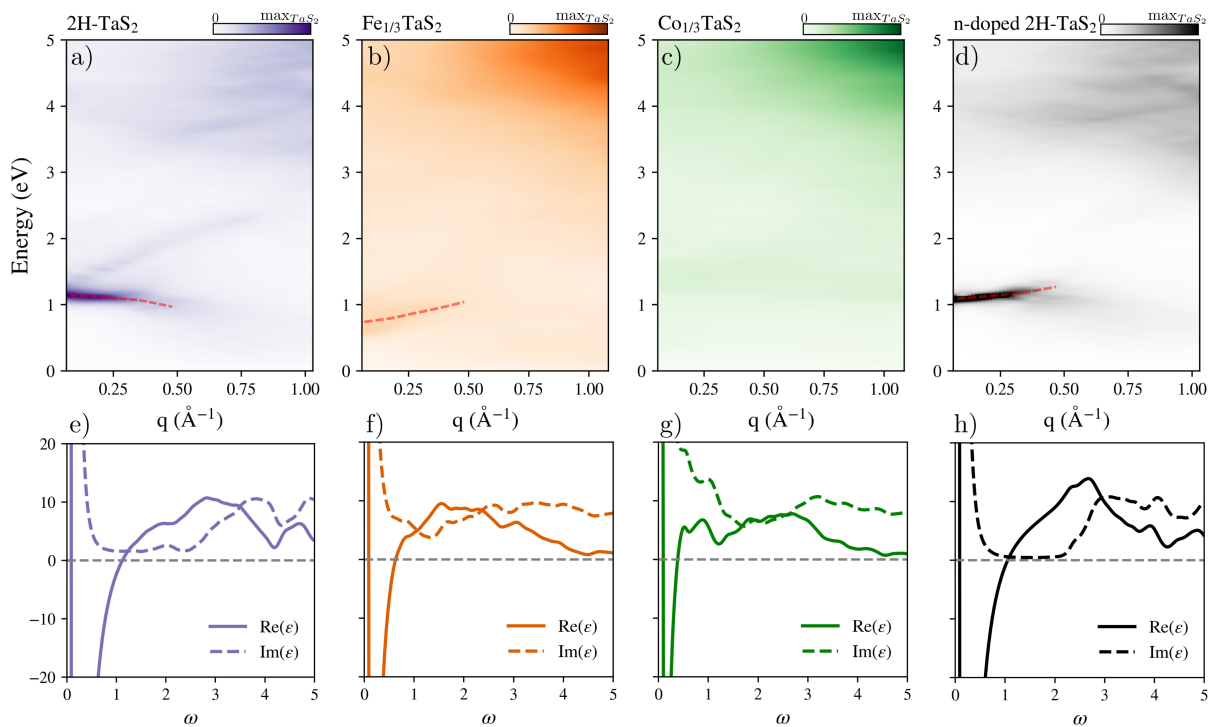

FIG. 3. a)-b)-c)-d) Energy-loss function heat map as a function of energy and momentum for 2H-TaS$_2$, Fe$_{1/3}$TaS$_2$, Co$_{1/3}$TaS$_2$ and electron doped 2H-TaS$_2$ . The heat map is calculated by linear interpolation of fixed momentum energy-loss function (see Methods for further details). In e)-f)-g)-h) we report real (solid line) and imaginary (dashed line) part of the dielectric function at $q$=0 for all the discussed cases.

TABLE I. Asymmetry parameter $\alpha$ for 2H-TaS$_2$, Fe$_{1/3}$TaS$_2$, and Co$_{1/3}$TaS$_2$ resulting from the fitting using Eq. (1) at different photon energies h$\nu$ (see Fig. 2).

|  | 300 eV | 700 eV | Al-K$_\alpha$ |
| --- | --- | --- | --- |
| 2H-TaS$_2$ | 0.17 | 0.41 | 0.75 |
| Fe$_{1/3}$TaS$_2$ | 0.04 | 0.03 | 0.12 |
| Co$_{1/3}$TaS$_2$ | 0.10 | 0.13 | 0.21 |

To further elucidate the microscopic origin of plasmon suppression and directly access its momentum dependence, we computed the electron energy-loss function (ELF) within linear-response theory using the random phase approximation (RPA) as implemented in GPAW [65–67] (see SI for details). In pristine 2H-TaS$_2$, the calculated plasmon dispersion reproduces the characteristic negative momentum dependence observed in TMDs [15, 18, 36, 37], reflecting the presence of isolated metallic bands and a highly nontrivial low-energy screening environment (Fig. 3a–b). Upon intercalation, this behavior is qualitatively altered. Both Fe and Co strongly suppress the plasmon peak in the ELF (Fig. 3c–f), with Co producing the most pronounced effect. This suppression is not merely a renormalization of the plasmon energy, but signals a breakdown of a well-defined collective mode. Microscopically, two concurrent mechanisms drive this evolution. First, intercalation reduces the bare plasma frequency, shifting the zeros of Re($\epsilon$) to lower energies (see Fig. 3e-f-g). Second, and more importantly, the intercalant-induced electronic states introduce a dense continuum of low-energy particle–hole excitations (Fig. 1 and Fig. 3e-f-g), which provide efficient decay channels for the plasmon. As a result, the collective excitation becomes strongly damped and progressively loses its coherence. This interpretation is directly confirmed by the calculated ELF: the plasmon mode of 2H-TaS$_2$, which originates from the isolated metallic bands [35], evolves into a broad, overdamped feature in the intercalated compounds due to the enhanced phase space for electronic excitations. Crucially, this behavior is fundamentally different from conventional electron doping. In the latter case, increasing carrier concentration preserves the coherence of the plasmon and typically sharpens the peak while modifying its dispersion, as shown in Fig. 3g–h and reported experimentally in Ref. [68]. Our results therefore identify intercalation as a distinct route to control collective excitations: rather than tuning carrier density alone, it reshapes the low-energy electronic structure through orbital hybridization and band reconstruction, effectively transforming a well-defined plasmon into a strongly damped excitation. This establishes a direct link between chemical complexity and dynamical screening, and highlights intercalation as a powerful tool to engineer plasmonic responses in layered quantum materials.

In conclusion, our combined spectroscopy and first-principles analysis demonstrates that Fe and Co intercalation provides a direct route to suppress and control the plasmon excitation characteristic of pristine 2H-TaS$_2$. This quenching arises not from simple electron doping but from structural reconstruction, orbital hybridization, and enhanced low-energy absorption channels that disrupt coherent screening. First-principle calculations confirm the loss of a sharp plasmon mode and a substantial modification of the peculiar negative dispersion in TMDs. These results establish intercalation as a chemically controlled route to engineer dynamical screening in layered metals highlighting a general mechanism for manipulating collective excitations in van der Waals quantum materials.

## I. DATA AVAILABILITY STATEMENT

All data that support the findings of this study are included within the article and supplementary materials.

## II. ACKNOWLEDGEMENTS

This work was funded by the European Union-NextGenerationEU under the Italian Ministry of University and Research (MUR) National Innovation Ecosystem Grant No. ECS00000041 VITALITY-CUP E13C22001060006. F.B. acknowledges funding from the National Recovery and Resilience Plan (NRRP), Mission 4, Component 2, Investment 1.1, funded by the European Union (NextGenerationEU), for the project "TOTEM" (CUP E53D23001710006 - Call for tender No. 104 published on 2.2.2022 and Grant Assignment Decree No. 957 adopted on 30/06/2023 by the Italian Ministry of University and Research (MUR)) and for the project "SHEEP" (CUP E53D23018380001 - Call for tender No. 1409 published on 14.9.2022 and Grant Assignment Decree No. 1381 adopted on 01/09/2023 by the Italian Ministry of Ministry of University and Research (MUR)). We acknowledge Elettra Sincrotrone Trieste for financial support under the SUI internal project and for providing access to its synchrotron radiation facilities (proposal number 20240507). This work has been partially funded through the project EUROFEL-ROADMAP ESFRI of MUR.

Supplemental material and supporting information for

# Plasmon Engineering in Intercalated vdW Materials

Luigi Camerano[a,b], Laura Martella [a], Lorenzo Battaglia [a], Federico Giannessi[a,b] Filippo Camilli [a] Polina M. Sheverdyaeva [c], Paolo Moras [c], Luca Lozzi [a], Luca Ottaviano[a,b], Gianni Profeta [a,b], Federico Bisti [a]

[a] Department of Physical and Chemical Sciences, University of L'Aquila, Via Vetoio 67100 L'Aquila, Italy
[b] CNR-SPIN L'Aquila, Via Vetoio, 67100 L'Aquila, Italy c/o Department of Physical and Chemical Sciences, University of L'Aquila, Via Vetoio, 67100 L'Aquila, Italy
[c] CNR-Istituto di Struttura della Materia (CNR-ISM), Strada Statale 14, km 163.5, 34149 Trieste, Italy

## CONTENTS

## I. EXPERIMENTAL METHODS

XPS core level spectra were acquired with PHI 1257 spectrometer with monochromatic Al K$_\alpha$ source with pass energy of 11.75 eV and a corresponding overall experimental resolution of 0.25 eV at room-temperature and base pressure 1x10$^{-9}$. LEED and photon-energy dependent core-level photoemission spectroscopy experiments were performed at the VUV-Photoemission beamline (Elettra, Trieste) at sample temperature T=20 K in ultrahigh vacuum (UHV) with a base pressure of 1x10$^{-10}$ mbar.

## II. THEORETICAL SIMULATIONS

Density functional theory calculations were performed using the Vienna ab-initio Simulation Package (VASP) [1, 2], using the generalized gradient approximation (GGA) in the Perdew-Burke-Ernzerhof (PBE) parametrization for the exchange-correlation functional [3], including SOC. Interactions between electrons and nuclei were described using the projector-augmented wave method. Energy thresholds for the self-consistent calculation was set to $10^{-6}$ eV and force threshold for geometry optimization $10^{-4}$ eV Å$^{-1}$. The Brillouin zone was sampled using an $8 \times 8 \times 4$ Gamma-centered Monkhorst-Pack grid. To account for the on-site electron-electron correlation on localized Fe-$d$ orbitals we used the GGA+U approach with an effective Hubbard term $U = 1.5$ eV. The 2H-TaS$_2$ lattice parameter are set to the experimental ones: $a = 3.31$ Å and $c = 12.07$ Å [4], while for the Fe$_{1/3}$TaS$_2$ $a = 5.737$ Å and $c = 12.28$ Å [5, 6] and Co$_{1/3}$TaS$_2$ $a = 5.725$ Å and $c = 11.878$ Å [7]. Due to the localization of Fe-$d$ orbitals, different metastable phases can be stabilized for Fe$_{1/3}$TaS$_2$. To stabilize large orbital moment phase we used a mixing parameter $\alpha_{mix} = 0.22$ and the occupation matrix control as implemented in VASP [8]. We notice that this phase is indeed the ground state of the system by comparing the total energy. The bare plasma frequencies are calculated by sampling the BZ using an $12 \times 12 \times 5$ Gamma-centered Monkhorst-Pack grid.

### A. ELF Computational details (GPAW)

To calculate the energy loss function (ELF), DFT calculations were performed using the projector augmented-wave (PAW) method as implemented by GPAW [9–11]. The exchange-correlation effects were treated using the PBE functional within the GGA approximation. Core-valence interactions were described using the PAW datasets supplied with GPAW. The experimentally reported crystal structures of TaS$_2$ and its Fe- and Co-intercalated derivatives were used without structural relaxation. For the self-consistent calculation of TaS$_2$, a plane-wave cutoff energy of 500 eV and a Monkhorst–Pack k-point mesh of $10 \times 10 \times 5$ were employed together with a Fermi–Dirac smearing of 0.05 eV. For the Fe- and Co-intercalated systems, a cutoff energy of 600 eV and a $6 \times 6 \times 3$ k-point mesh were used, together with a 0.2 eV Fermi-Dirac smearing. The convergence criteria and number of bands were kept at the default values implemented in GPAW. Spin-polarized calculations were performed for the intercalated systems. An initial magnetic moment of $2\,\mu_\text{B}$ was assigned to the Fe and Co atoms, finally converging to $3\,\mu_\text{B}$ for Fe and $1.3\,\mu_\text{B}$ for Co. For the Fe-intercalated system a ferromagnetic ordering was imposed while for Co-intercalated system an A-type antiferromagnetic configuration was imposed as the initial magnetic ordering. The final magnetic states were obtained self-consistently within the SCF procedure and used for subsequent calculations. For the intercalated systems, electronic convergence was also facilitated using the density mixing scheme MixerSum, with parameters $\beta = 0.1$, nmaxold = 5 and weight = 50.0). For the Fe-intercalated system, on-site Coulomb interactions were included within the DFT+$U$ formalism with $U = 1.5$ eV applied to the Fe $d$ states. Non-self-consistent calculations were next carried out using the converged ground-state density to increase the Brillouin zone sampling. The k-point meshes were refined to $30 \times 30 \times 15$ for TasS$_2$ and $20 \times 20 \times 10$ for the intercalated systems. The dielectric function and energy loss function were calculated within the linear response formalism using the random phase approximation (RPA) as implemented in GPAW. A local field effect cutoff of 100 eV was included in the response function calculation, together with a broadening parameter of $\eta = 25 \cdot 10^{-3}$ eV. The frequencies ranged from 0.01 to 5 eV. The macroscopic dielectric function was computed along the $x$-direction using the same parameters. The momentum-resolved ELF was evaluated along the Γ–M direction for momentum transfers ranging from $0.06 \, \text{Å}^{-1}$ to $1.07 \, \text{Å}^{-1}$.

## III. PLASMA FREQUENCIES AND DIELECTRIC FUNCTION

In this section we report the calculated bare plasma frequencies for the different compounds in Table S1 and the real Re($\epsilon$) and imaginary Im($\epsilon$) part of the dielectric function. We note the in-plane bare plasma frequency $\omega_{p,\parallel}$ decreases upon intercalation while the out-of-plane bare plasma frequency $\omega_{p,\perp}$ increases. This is pointing to an out-of-plane hybridization induced by intercalation as confirmed in ARPES [12]. The decreasing of in-plane plasma frequency directly influence the zero of the real part of the dielectric function $\text{Re}(\epsilon(\omega))$ which determine the plasmon resonance condition as reported in Fig.3 of the main text. As discussed in the main manuscript, the suppression of the plasmon is due to the combined effect of a reduction of the bare plasma frequency and the introduction of additional low-energy states upon intercalation resulting in incresed $\text{Im}(\epsilon(\omega))$ at low energy.

TABLE S1. In-plane ($\parallel$) and out-of-plane ($\perp$) bare plasma frequencies ($\omega_p$) in eV for $TaS_2$, $Fe_{1/3}TaS_2$, and $Co_{1/3}TaS_2$.

|  | $\omega_{p,\parallel}$ | $\omega_{p,\perp}$ |
| --- | --- | --- |
| $TaS_2$ | 3.24 | 0.28 |
| $Fe_{1/3}TaS_2$ | 2.56 | 1.11 |
| $Co_{1/3}TaS_2$ | 1.88 | 0.97 |

## IV. FERMI LEVEL ALIGNMENT

In Fig. S1 we report the Fermi-edge fits obtained using a logistic function, while Fig. S2 shows the valence band, Ta-$4f$, and S-$2p$ spectra for all investigated compounds, aligned to the corresponding fitted Fermi level. As discussed in the main text, Fig. S2a shows a doping of the valence band which is absent in the Ta-$4f$ core level Fig. S2c, signaling efficient metallic screening.

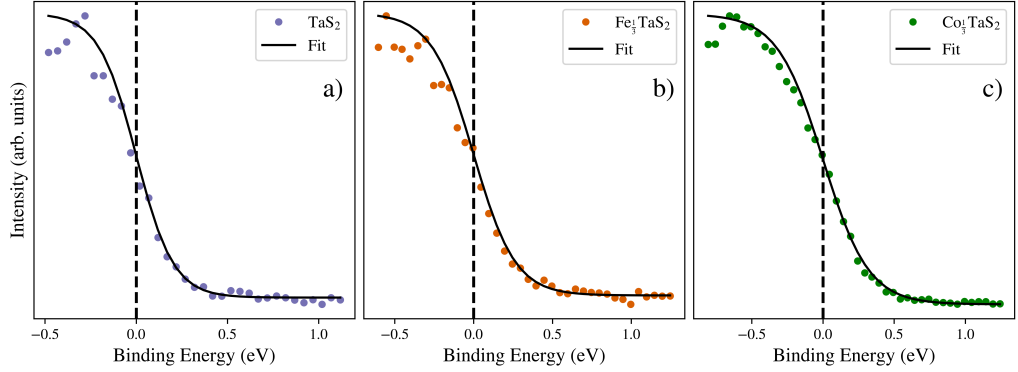

FIG. S1. Fitted Fermi level edge for 2H-TaS$_2$ (a), Fe$_{1/3}$TaS$_2$ (b) and Co$_{1/3}$TaS$_2$ (c).

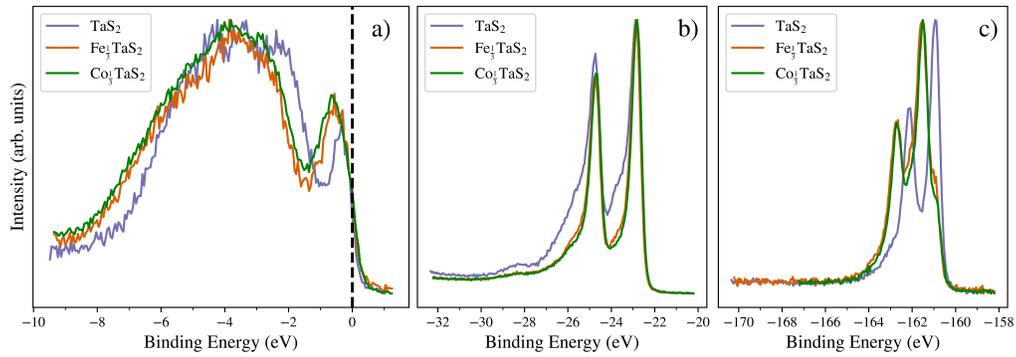

FIG. S2. Valence band (a), Ta-$4f$ (b) and S-$2p$ (c) core-level spectra acquired at room temperature using Al-K$_\alpha$ source (h$\nu$ = 1486.7 eV) for 2H-TaS$_2$, Fe$_{1/3}$TaS$_2$ and Co$_{1/3}$TaS$_2$. All spectra are referenced to the respective Fermi energies.

## V. XPS DATA ANALYSIS

In this section, we introduce a phenomenological lineshape that accounts for both JDoS effects and extrinsic losses in the core levels of 2H-TaS$_2$ and its intercalated counterparts. In Fig. S3, we demonstrate that a simple Voigt doublet is insufficient to capture the loss features and the asymmetry arising from the electronic JDoS. To address this, we adopt a skew normal distribution (SKND), which provides an improved description of the lineshape in these systems. The effectiveness of this approach is illustrated in Figs. S4 and S5, where the experimental spectra are accurately reproduced. Additionally, in Fig. S4 we identify an extra component with a main peak centered at 26 eV, which we attribute to the formation of Ta$_2$O$_5$ (Ta$^{5+}$) [13]. Finally, Fig. S6 shows the S-2$p$ core-level spectra of Co$_{1/3}$TaS$_2$ as a function of the emission angle. The ratio between bulk and surface components, reported in Table S2, further supports the surface origin of the additional spectral feature.

The analytic expression of the SKND is:

$$N(E; \mu, \sigma, \alpha) = \frac{2\mathrm{A}}{\sigma} \, \phi\left(\frac{E-\mu}{\sigma}\right) \Phi\left(\alpha \cdot \frac{E-\mu}{\sigma}\right) \tag{1}$$

with:

- $\phi(E; \mu, \sigma) = \frac{1}{\sqrt{2\pi\sigma^2}} e^{-\frac{(E-\mu)^2}{2\sigma^2}}$ is the standard normal distribution,

- $\Phi(E; \mu, \sigma) = \frac{1}{2}\left[1 + \mathrm{erf}\left(\frac{E-\mu}{\sqrt{2}\sigma}\right)\right]$ is the associated partition function,

- $\alpha \in \mathbb{R}$ is the asymmetry parameter.

- A is the amplitude.

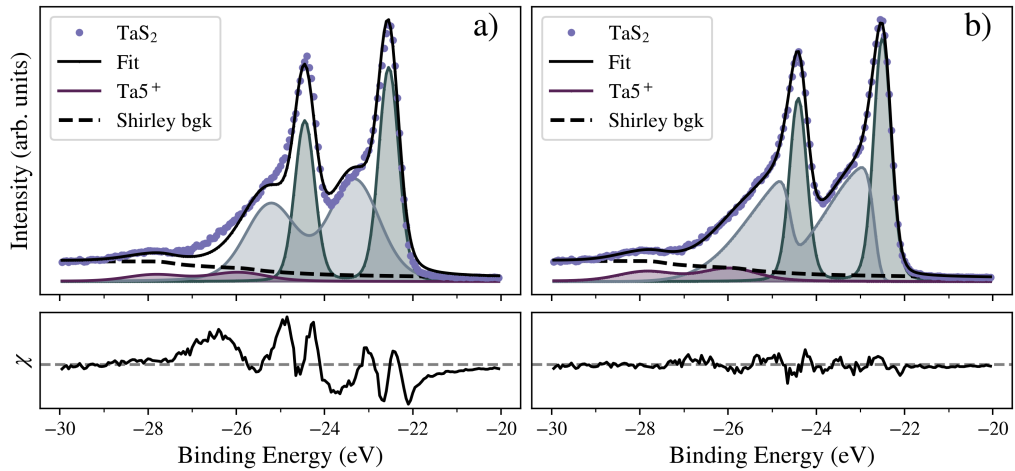

FIG. S3. Ta-4$f$ core-level spectra of 2H-TaS$_2$ acquired at room temperature using an Al-K$_\alpha$ source ($h\nu = 1486.7$ eV). Top panels a) and b) show the experimental data and corresponding fits, while the bottom panels display the residuals $\chi$ between the data and the fitting functions. All plots share the same energy scale. a) The fit consists of three Voigt doublets: the main emission, an additional feature accounting for the extrinsic plasmon, and a Ta$^{5+}$ doublet due to surface oxidation. b) The fit utilizes Voigt functions for the main and oxidation doublets, and a Skew Normal Distribution (SKND) to describe the both the extrinsic plasmon and asymmetry of the peak due to electronic JDoS effects. In both panels, the black dashed line represents the Shirley background and all the doublets share the same ratio and peaks distance.

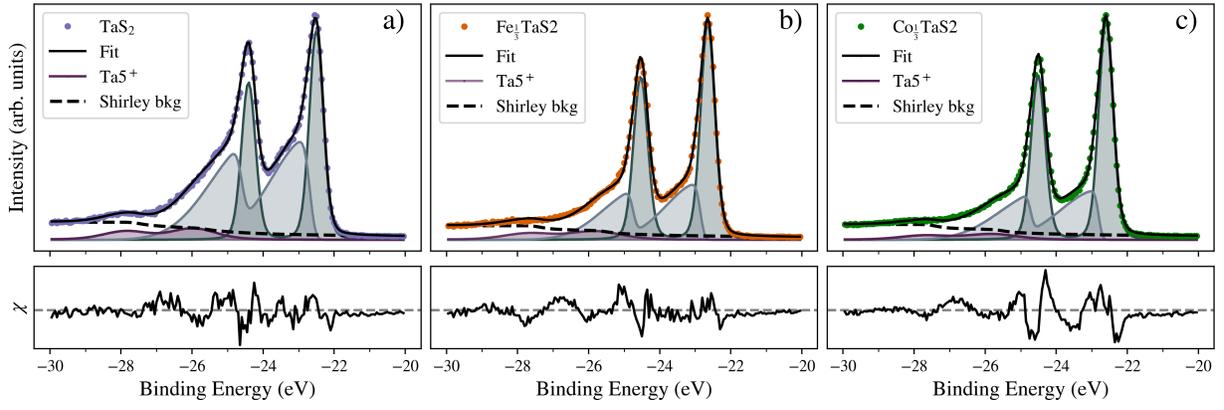

FIG. S4. Ta-4f core-level spectra of 2H-TaS$_2$ (a), Fe$_{1/3}$TaS$_2$ and Co$_{1/3}$TaS$_2$ acquired at room temperature using an Al-K$_\alpha$ source ($h\nu = 1486.7$ eV). Top panels show the experimental data and corresponding fits, which uses SKND for the extrinsic plasmon. Bottom panels display the residuals $\chi$ between the data and the fitting functions. All plots share the same energy scale. The black dashed line represent the Shirley background. All the doublets share the same ratio and peaks distance.

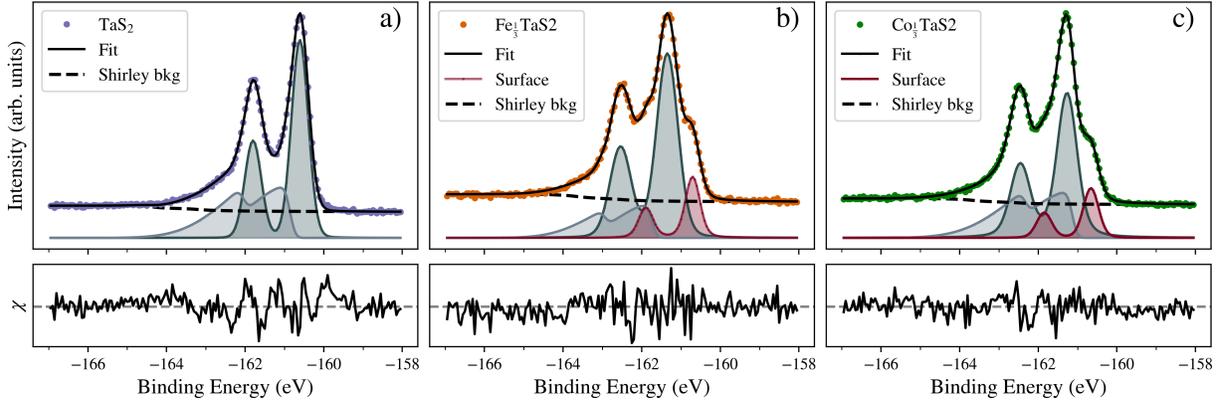

FIG. S5. S-2p core-level spectra acquired at room temperature using an Al-K$_\alpha$ source ($h\nu = 1486.7$ eV). a) 2H-TaS$_2$ spectrum, the fit employs a Voigt doublet for the main emission and a SKND doublet for the extrinsic plasmon. b) 2H-Fe$_{1/3}$TaS$_2$ spectrum, the fit consists in 2 Voigt doublets for main emission and surface S, and a SKND doublet. c) Same quantities for 2H-Co$_{1/3}$TaS$_2$. All the doublets share the same ratio and peaks distance. The black dashed line represent the Shirley background. Bottom panels display the difference $\chi$ between the data and the fitting functions. All plots share the same energy scale.

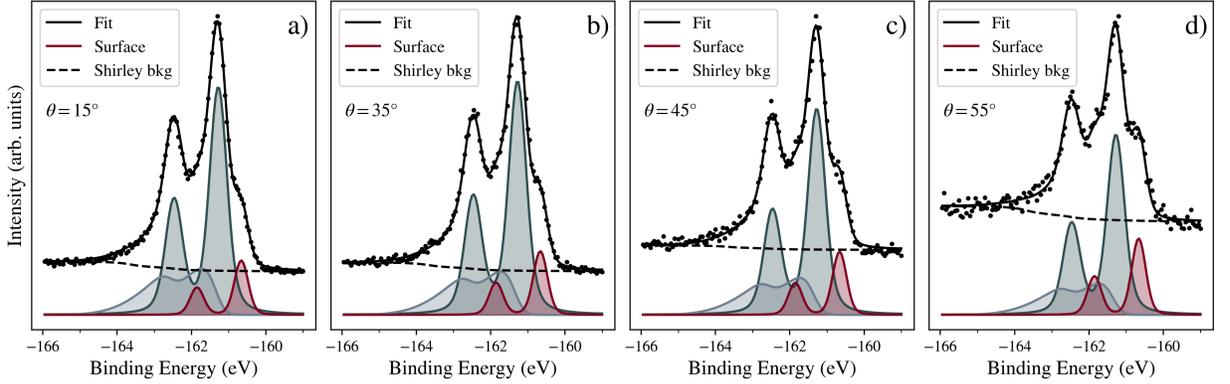

FIG. S6. S-2p core-level spectra of 2H-Co$_{1/3}$TaS$_2$ acquired at room temperature using an Al-K$_\alpha$ source ($h\nu = 1486.7$ eV), varying the angle between the incident radiation and the sample. In each panel the fit consists in 2 Voigt doublets for main emission, surface S and a SKND doublet. The lineshape parameters are fixed to the values derived from the fit in Fig. S5 c), leaving only the peak intensities as free parameters. The black dashed line represent the Shirley background. All plots share the same energy scale.

TABLE S2. Ratio between the area of the main emission peak and the peak of surface S

|  | 15° | 35° | 45° | 55° |
|---|---|---|---|---|
| A$_{\text{surface}}$/A$_{\text{main}}$ | 0.239 | 0.273 | 0.307 | 0.424 |

## A. Discussion on possible contamination

In Ref. [14], the Ta-4$f$ core-level spectrum was modeled using multiple components, two of which were attributed to contamination effects, with the absence of plasmon-related losses ascribed to the low crystallinity of the samples. Our measurements are performed on high-quality single crystals cleaved *in situ* under ultra-high vacuum conditions, ensuring pristine and well-ordered surfaces. The resulting spectra, particularly those acquired at the synchrotron, exhibit sharp and well-defined Ta-4$f$ and S-2$p$ core-level features, indicative of a uniform chemical environment across the sample. Moreover, contamination-related contributions are expected to be surface-sensitive and therefore enhanced at lower photon energies, where the probing depth is reduced. This behavior is not observed in our data. On the contrary, the ∼1 eV feature in the Ta-4$f$ spectra increases in intensity with increasing photon energy, consistent with a bulk-sensitive origin. This trend follows the increase of the inelastic mean free path, which enhances the probability of photoelectrons undergoing inelastic scattering processes before escaping the sample, supporting its assignment to an extrinsic loss feature that are enanched with photon energy as is shown in Ref. [15] where Hard-X-rays XPS were used to study extrinsic plasmon excitation in aluminum. Further evidence against contamination effects is provided by the S-$2p$ spectra of pristine 2H-TaS$_2$, which display a single, well-defined main component with no additional features across all measured photon energies. Indeed, possible sulfur oxidation states will appear at lower binding energy (in the range (-166,-170) eV) as reported in Ref. [16]. Additionally, no signatures of Ta oxides formation are observed in the Ta-$4f$ lineshape in our synchrotron measurements, in contrast to laboratory XPS data where surface oxidation can be detected. Finally, the additional shoulder in S-$2p$ core-level fitted with asymmetric lineshape in Fig. S5, that is more evident in 2H-TaS$_2$ spectra, further confirm the presence of losses in core-level spectra. Taken together, these observations demonstrate that the additional spectral weight observed in the core levels originates from electronic excitations of the material, manifesting as extrinsic plasmon-loss processes during photoemission, rather than from contamination or sample inhomogeneity.